 \documentclass[twocolumn,prl,aps,psfig,showpacs,preprintnumbers,superscriptaddress]{revtex4}
\usepackage{graphicx}
\usepackage{epstopdf}
\usepackage{float}
\usepackage{color}
\usepackage{amsmath}
\usepackage{bm}

\begin{document}

\title{Magnetic properties of the itinerant A-type antiferromagnet CaCo$_2$P$_2$ studied by $^{59}$Co and $^{31}$P NMR}
\author{N. Higa$\footnote[1]{Present address: Advanced Science Research Center, Japan Atomic Energy Agency, Tokai, Ibaraki, 319-1195, Japan}$}
\affiliation{Ames Laboratory, U.S. DOE, and Department of Physics and Astronomy, Iowa State University, Ames, Iowa 50011, USA}
\affiliation{Department of Physics and Earth Sciences, Faculty of Science, University of the Ryukyus, Okinawa 903-0213, Japan}
\author{Q.-P. Ding}
\affiliation{Ames Laboratory, U.S. DOE, and Department of Physics and Astronomy, Iowa State University, Ames, Iowa 50011, USA}
\author{A. Teruya}
\affiliation{Department of Physics and Earth Sciences, Faculty of Science, University of the Ryukyus, Okinawa 903-0213, Japan}
\author{M. Yogi}
\affiliation{Department of Physics and Earth Sciences, Faculty of Science, University of the Ryukyus, Okinawa 903-0213, Japan}
\author{M. Hedo}
\affiliation{Department of Physics and Earth Sciences, Faculty of Science, University of the Ryukyus, Okinawa 903-0213, Japan}
\author{T. Nakama}
\affiliation{Department of Physics and Earth Sciences, Faculty of Science, University of the Ryukyus, Okinawa 903-0213, Japan}
\author{Y. \=Onuki}
\affiliation{Department of Physics and Earth Sciences, Faculty of Science, University of the Ryukyus, Okinawa 903-0213, Japan}
\author{Y. Furukawa}
\affiliation{Ames Laboratory, U.S. DOE, and Department of Physics and Astronomy, Iowa State University, Ames, Iowa 50011, USA}

\date{\today}

\begin{abstract} 
    $^{59}$Co and $^{31}$P nuclear magnetic resonance (NMR) measurements in external magnetic and zero magnetic fields have been performed to investigate the magnetic properties of the A-type antiferromagnetic (AFM)  CaCo$_2$P$_2$. 
    NMR data, especially, the nuclear spin lattice relaxation rates 1/$T_1$ exhibiting  a clear peak, provide clear evidence for the AFM transition at a N\'eel temperature of $T_{\rm N}\sim$110~K. 
    The magnetic fluctuations in the paramagnetic state were  found to be three-dimensional ferromagnetic, suggesting ferromagnetic interaction between Co spins in the ${\it ab}$ plane characterize the spin correlations in the paramagnetic state.   
     In the AFM state below $T_{\rm N}$, we have observed $^{59}$Co and $^{31}$P NMR signals under zero magnetic field.
     From $^{59}$Co NMR data, the ordered magnetic moments of Co are found to be in $ab$ plane and are estimated to be 0.35 $\mu_{\rm B}$ at  4.2  K. 
     Furthermore, the external field dependence of $^{59}$Co NMR spectrum in the AFM state suggests a very weak  magnetic  anisotropy of the Co ions and also provides microscopic evidence of canting the Co ordered moments along the external magnetic field directions.  
     The magnetic state of the Co ions in CaCo$_2$P$_2$  is well explained by the local moment picture in the AFM state, although the system is metallic as seen by  $1/T_1T$ = constant behavior.

\end{abstract}

\maketitle

  \section{I.  Introduction}

  Since the discovery of high $T_{\rm c}$ superconductivity (SC) in iron pnictides \cite{Kamihara2008},  the interplay between spin fluctuations and the unconventional nature of SC has received wide interest.  
    In most of the iron pnictide superconductors, by lowering temperature, the crystal structure changes from high-temperature tetragonal ($C_4$ symmetry)  to low-temperature orthorhombic ($C_2$ symmetry) at, or just above, a system-dependent N\'eel temperature $T_{\rm N}$, below which long-range stripe-type antiferromagnetic (AFM) order emerges \cite{Canfield2010,Johnston2010,Stewart2011}.
   SC in these compounds emerges upon suppression of both the structural (or nematic) and magnetic transitions by carrier doping and/or the application of pressure.
   The interplay between SC and  AFM spin fluctuations and nematic fluctuations is still an open question. 
    In addition,  ferromagnetic (FM) correlations were also pointed out  to play an important role in the iron-based superconductors  \cite{Johnston2010,Nakai2008,PaulPRB, PaulPRL, Cui2017}.
    
      Recently, new magnetic states with $C_4$ symmetry have attracted much attention \cite{Hoyer2016}.  
      The so-called charge-spin density wave (CSDW) has been demonstrated to be realized in Sr$_{1-x}$Na$_x$Fe$_2$As$_2$ \cite{Allred2016}, and likely occurs in Ba(Fe$_{1-x}$Mn$_x$)$_2$As$_2$, Ba$_{1-x}$Na$_x$Fe$_2$As$_2$, and Ba$_{1-x}$K$_x$Fe$_2$As$_2$ as well  \cite{Kim2010,Avci2014,Wang2016,Hassinger2012,Bohmer2015,Allred2015,Hassinger2016}.
      In addition, a new magnetic state called $``$hedgehog" spin vortex crystal with $C_4$ symmetry has been identified in the electron doped 1144-type iron pnictide superconductors CaK(Fe$_{1-x}M_{x}$)$_4$As$_4$ ($M$ = Co or Ni) \cite{Meier20172,DingPRB2017,Kreyssig2018}.
      In contrast, nonmagnetic state has been observed in the collapsed tetragonal phase with $C_4$ symmetry in such as CaFe$_2$As$_2$ \cite{Kreyssig2008,Pratt2009,Prokes2010,Ran2011} where AFM fluctuations are also revealed to be completely suppressed \cite{Kawasaki2010,Soh2013,Furukawa2014}.

    Such discoveries naturally triggered the investigation of the magnetic properties of a wide variety of other layered pnictides. 
    Among them, the Co compounds have been found to show a rich variety of magnetic properties with different crystal structure.
    Tetragonal SrCo$_2$P$_2$ with $C_4$ symmetry shows no magnetic ordering and is an exchange enhanced Pauli paramagnet with a dominantly ferromagnetic interactions \cite{Moresen1998, Jia2009, Imai2014, Imai2015PP, Imai2017}.
    A metamagnetic transition from the Pauli paramagnetic to the ferromagnetic state at a high magnetic field of 60 T has been reported \cite{Imai2014}.
    LaCo$_2$P$_2$ with the tetragonal structure is known to be an itinerant ferromagnet with a Curie temperature of $T_{\rm C}$ = 130 K and a saturated Co moment of 0.4 $\mu_{\rm B}$ \cite{Reehuis1994,Imai2015PRB, Teruya_LaCo2P2}.

     In CaCo$_2$P$_2$ with the collapsed tetragonal structure, on the other hand, an A-type AFM  state  has been reported below a N\'eel temperature of 110 K, in which the Co moments are ferromagnetically aligned in the $ab$ plane and the moments adjacent along the $c$ axis are aligned antiferromagnetically \cite{Reehuis1998,Imai2017}.
    This is in contrast to the nonmagnetic state in the case of the Fe pnictides with the collapsed tetragonal structure.   
    From the neutron diffraction (ND) measurements on powder samples of CaCo$_2$P$_2$,  the ordered Co moments are estimated to be 0.32 $\mu_{\rm B}$ \cite{Reehuis1998}.
    The magnetic susceptibility increases by lowering temperature and exhibits  a small kink at $T_{\rm N}$ \cite{Baumbach2014PRB,Teruya_LaCo2P2}.
    Even below $T_{\rm N}$, the magnetic susceptibility keeps increasing and shows a broad maximum at $T$$^*$ $\sim$ 32-36 K whose origin is not well understood yet  \cite{Baumbach2014PRB,Teruya_LaCo2P2}.
    Teruya {\it et~al}. suggested that the anomaly at $T^*$ is intrinsic and relates to the metamagnetic-like behavior observed in the magnetization data \cite{Teruya_LaCo2P2}.

     In this paper, we report a comprehensive study of $^{59}$Co and $^{31}$P nuclear magnetic resonance (NMR) measurements in external and zero magnetic fields in the paramagnetic (PM) and the AFM states of CaCo$_2$P$_2$.  
     NMR shift  and nuclear spin-lattice relaxation time data clearly show the dominant ferromagnetic spin fluctuations in the paramagnetic state, which is consistent with 
the previous  $^{31}$P NMR measurements up to 200 K reported by Imai {\it et~al}. \cite{Imai2017}. 
    Our experimental data up to 300 K  clearly evidence that the ferromagnetic fluctuations are isotropic and of three dimensional in nature.
    Observation of $^{59}$Co NMR signals in zero magnetic field is direct evidence of the AFM ordering below $T_{\rm N}$ = 110 K.
    The Co ordered moments are revealed to be in the $ab$ plane and are estimated to be 0.35 $\mu_{\rm B}$, consistent with the neutron diffraction measurements \cite{Reehuis1998}.
     No obvious change in the magnitude of the Co ordered moments at $T^*$ indicates that the relative orientations of Co ordered moments between the layers change.
   The metamagnetic behavior observed in magnetization data is also explained by the small change in the canting angle of the Co ordered moments. 

 \section{II. Experimental}
 
     A single crystal of CaCo$_2$P$_2$ was grown by Sn-flux method. 
    Details of the sample preparation are described elsewhere~\cite{Teruya_LaCo2P2}. 
    NMR measurements were performed on $^{31}$P ($I$ = 1/2, $\gamma_{\rm n}/2\pi$ = 17.235~MHz/T) and $^{59}$Co ($I$ = 7/2, $\gamma_{\rm n}/2\pi$ = 10.03~MHz/T) by  using a homemade phase-coherent spin-echo pulse spectrometer. 
    The $^{59}$Co and  $^{31}$P-NMR spectra were obtained by sweeping external magnetic field $H$ at fixed frequencies  while NMR spectra in zero or small magnetic fields were measured in steps of frequency by measuring the intensity of the Hahn spin echo.  
    The $^{31}$P  spin-lattice relaxation rates 1/$T_1$ were measured by saturation recovery method. 
    The nuclear magnetization recovery was found to follow a single exponential function at the measured temperature region.

\section{III. Results and discussion}
\subsection{A.   $^{59}$Co NMR in the paramagnetic state }

\begin{figure}[tb]
	\centering
	\includegraphics[width=8.5cm]{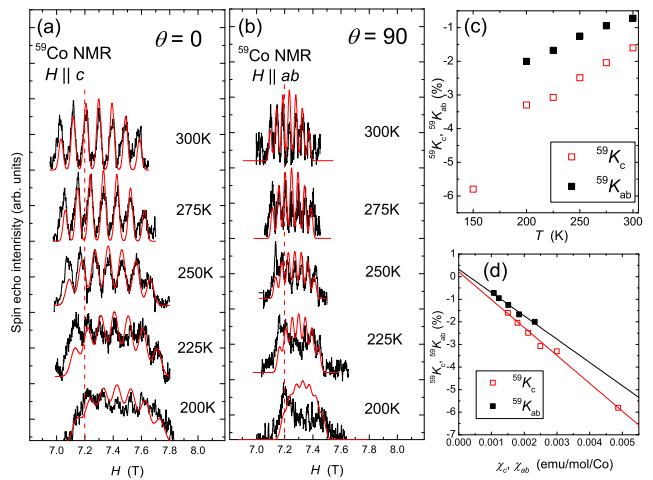} 
	\caption{(Color online)  Field-swept $^{59}$Co-NMR spectra at a frequency of $f$ = 72.2~MHz at various temperatures for magnetic fields (a) $H\parallel c$ axis and (b) $H\parallel ab$ plane. The red solid curves are simulated spectra with a nearly temperature independent of  $\nu_{\rm Q}$ = 0.9~MHz. The vertical dashed red  lines correspond to the Lamor magnetic field. 
(c)~$T$ dependence of the $^{59}$Co-NMR shifts $^{59}K_{ab}$ and $^{59}K_{c}$ above $T_{\rm N}$. 
(d)~$K-\chi$ plots for $H\parallel ab$ plane (black) and $H\parallel c$ axis (red), where we used $\chi$ data reported in Ref.~\onlinecite{Teruya_LaCo2P2}. 
The solid lines are fitting results. }
	\label{fig:Fig1}
\end{figure}

     Figures ~\ref{fig:Fig1}(a) and \ref{fig:Fig1}(b) show the temperature ($T$) dependence of field-swept $^{59}$Co-NMR spectra for magnetic fields parallel to  the $c$ axis ($H\parallel c$) and to the $ab$ plane ($H \parallel ab$), respectively. 
       The typical spectrum for a nucleus with spin $I=7/2$ with Zeeman and quadrupolar interactions can be described by a nuclear spin Hamiltonian \cite{Slichter_book} 
\begin{eqnarray}
\centering
{\cal H} &=& -\gamma_{\rm n}\hbar(1+K) {\bf H} \cdot {\bf I} + \frac{h\nu_{\rm Q}}{6}(3I_Z^2-I^2),
\label{eq:1}
\end{eqnarray} 
where $H$ is external field, $h$ is Planck's constant and $K$ represents the NMR shift. 
   The nuclear  quadrupole frequency for $I=7/2$ nuclei is given by $\nu_{\rm Q} = eQV_{\rm ZZ}/14h$, where $Q$ is the nuclear quadrupole moment and $V_{\rm ZZ}$ is the electric field gradient (EFG) at the nuclear site.
    In first order perturbation theory, when the Zeeman interaction is much greater than the quadrupole interaction, one has the nuclear energy level for $I$ = 7/2 
\begin{eqnarray}
\centering
E_m  = -\gamma_{\rm n}\hbar(1+K)Hm - \frac{h\nu_{\rm Q}}{12}(3\cos^2\theta -1 )(3m^2-\frac{63}{4}), 
\label{eq:1}
\end{eqnarray}
where $\theta$ is the angle between the external magnetic field and the principal axis of the EFG. 
    Thus $^{59}$Co  NMR spectrum is composed of a central transition line  and  three pairs of satellite lines shifted from the central transition line by $\pm\frac{1}{2} \nu_{\rm Q}(3\cos^2\theta -1 )$ (for the transitions of $m$ = 3/2 $\leftrightarrow$ 1/2 and -3/2 $\leftrightarrow$ -1/2), $\pm \nu_{\rm Q}(3\cos^2\theta -1 )$ (for $m$ = 5/2 $\leftrightarrow$ 3/2 and -5/2 $\leftrightarrow$ -3/2), and $\pm\frac{3}{2}  \nu_{\rm Q}(3\cos^2\theta -1 )$ (for $m$ = 7/2 $\leftrightarrow$ 5/2 and -7/2 $\leftrightarrow$ -5/2).
   It is noted that the spacing between the lines of the spectrum  for $\theta$ = 0  is twice of that for $\theta$ = $\pi$/2, producing the spectrum for  $\theta$ = 0 almost two times wider than for $\theta$ = $\pi$/2.
   The observed $^{59}$Co NMR spectra show the clear seven distinct lines above 250~K as expected, and were well reproduced by simulated spectra (red lines) from the simple Hamiltonian with $\nu_{\rm Q}$ = 0.9~MHz which is found to be independent of temperature at least above $\sim$200 K.
       From the spectrum analysis where  $\theta$ is found to be 0 and $\pi/$2 for $H \parallel c$ and $H \parallel ab$, respectively,  it is clear that the principal axis of the EFG at the Co site is along the $c$ axis. 
     This is consistent with the local tetragonal symmetry (4$mm$) of the Co ions in the crystal. 

      As shown in the Fig.~\ref{fig:Fig1}, with decreasing temperatures, each line becomes broader due to inhomogeneous magnetic broadening and the spectra show no clear feature of the quadrupolar split lines below $\sim$ 200 K. 
     At the same time, nuclear spin-spin relaxation time $T_2$ becomes shorter at low temperatures. 
     Those make NMR spectrum measurements difficult below $\sim$ 200 K.  
     It is noted that the two peak structure observed at 225 K  for $H\parallel ab$  is due to the shorter $T_2$ of the central transition lines than the satellite lines, making the difference in the signal intensity between the experimental and the simulated results around the center of the spectra. 
     Similar double peak structure has been observed in $^{51}$V NMR spectra in the AFM compound YVO$_3$ \cite{YVO3} and also in the kagome staircase compound Ni$_3$V$_2$O$_8$ \cite{Vasily2010}.

   Figure~\ref{fig:Fig1}(c) shows  the $T$ dependence of the $^{59}$Co-NMR shift for $H \parallel ab$ plane  ($^{59}K_{ab}$) and $H\parallel c$  axis ($^{59}K_{c}$)  determined from the simulated spectra, where both $^{59}K_{ab}$  and $^{59}K_{c}$ decrease on lowering temperature.  
   The NMR shift consists of temperature  dependent spin shift $K_{\rm s}(T)$ and $T$ independent orbital shift $K_{\rm 0}$; $K(T)$ =$K_{\rm s}(T)$ + $K_{\rm 0}$ where $K_{\rm s}(T)$ is proportional to the spin part of magnetic susceptibility  $\chi_{\rm s}$($T$) via hyperfine coupling constant $A$, $K_{\rm s}(T)$  = 
$\frac{A\chi_{\rm s}(T)}{N_{\rm A}}$.  
   Here  $N_{\rm A}$ is Avogadro's number.   
   The hyperfine coupling constants are estimated to be $A_{ab}^{\rm Co}$ = (-57.6 $\pm$ 4.2)~kOe/$\mu_{\rm B}$/Co and $A_{c}^{\rm Co}$ = (-68.9$ \pm$ 3.5)~kOe/$\mu_{\rm B}$/Co from the slopes of the so-called $K$-$\chi$ plot shown in Fig.~\ref{fig:Fig1}(d).
  These values are smaller than a typical value $A$ = -105~kOe/$\mu_{\rm B}$ for 3$d$ electron core polarization \cite{Abragam1955}.
      Similar reductions in the hyperfine coupling constant have been observed in several Co compounds \cite{Tsuda1968,Fukai1996,Roy2013,PaulPRB}.
    As discussed in Refs. \onlinecite{Tsuda1968,Fukai1996,Roy2013}, the small values could be due to anisotropic and positive orbital and/or dipolar hyperfine coupling contributions which cancel a part of the negative core polarization hyperfine field.    
    $K_{\rm 0}$ are estimated to be $^{59}K_{0,ab}$  = 0.33 $\%$ and $^{59}K_{0,c}$  = 0.21 $\%$,  for $H \parallel ab$ and $H \parallel c$, respectively.

\subsection{B.  $^{31}$P NMR spectrum}

\begin{figure}[tb]
	\centering
	\includegraphics[width=8.5cm]{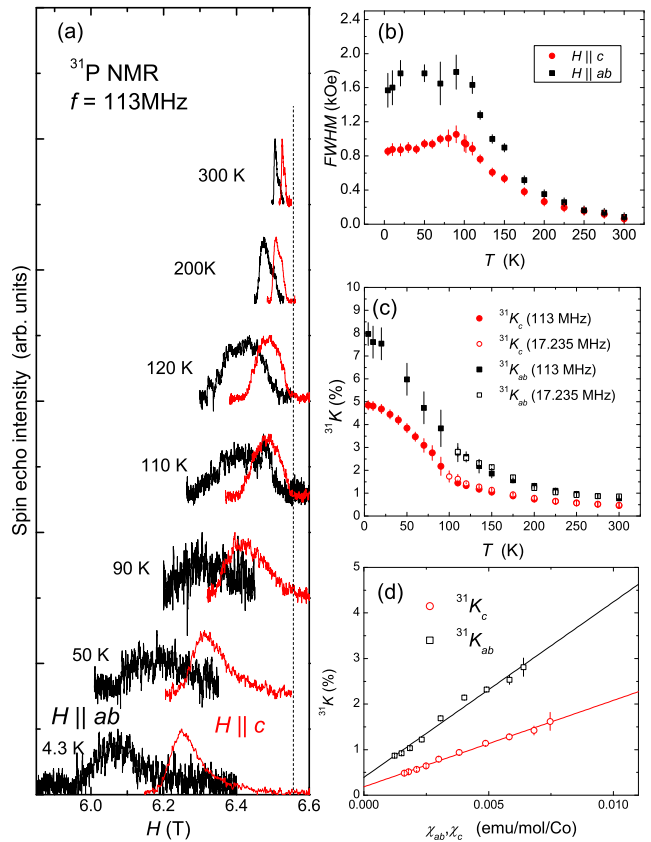} 
	\caption{(Color online) Field-swept $^{31}$P-NMR spectra at a frequency of $f$ = 113~MHz at various temperatures for magnetic fields parallel to the $c$ axis (red) and to the $ab$ plane (black). The vertical dashed black line represents the zero-shift position ($^{31}K$ = 0).
    (b) $T$ dependence of $FWHM$ for $H\parallel c$ and $H\parallel ab$. 
    (c) $T$ dependence of the $^{31}$P-NMR shifts $^{31}K_{ab}$ (black) and $^{31}K_{c}$ (red)  at $f$ = 113~MHz (closed symbols) and 17.235~MHz (open symbols).	
    (d) $K$-$\chi$ plots for $H \parallel ab$ plane (black) and $H \parallel c$ axis (red) above $T_{\rm N}$. Here we used $K$ data measured at 17.235 MHz and $\chi$ data from Ref.~\onlinecite{Teruya_LaCo2P2}. The solid lines are fitting results. }
	\label{fig:Fig2}
\end{figure}

      Typical field-swept $^{31}$P-NMR spectra at a  frequency of $f$ = 113~MHz at several temperatures for $H \parallel ab$ plane (black lines) and $H \parallel  c$ axis (red lines) are shown in  Fig.~\ref{fig:Fig2}(a). 
      Although the NMR  line is relatively sharp at high temperatures, the spectra broaden with decreasing temperatures and become  asymmetric.   
      Since $^{31}$P nucleus has  $I$ = 1/2, the asymmetry of the line cannot be due to quadrupole interaction. 
       In addition, we used a single crystal for our measurements.  
       Therefore  the asymmetric shape indicates the distribution of the local internal field at the P sites, suggesting the inhomogeneity of Co magnetic moments.  
      As shown in  Fig.~\ref{fig:Fig2} (b), the values of full-width at half maximum ($FWHM$) of the spectra for both magnetic field directions increase with decreasing temperature, but both $FWHM$s are nearly constant below $T_{\rm N}$.
        Similar results have been reported previously \cite{Imai2017}.
       Those results indicate that the inhomogeneity of the Co magnetic moments does not show any increase although the magnetic susceptibility keeps increasing down to $T^*$ $\sim$ 32-36 K.

     Figure~\ref{fig:Fig2}(c) represents the  $T$ dependence of NMR shift for both magnetic fields $H \parallel ab$ plane ($^{31}K_{ab}$) and $H \parallel c$ axis ($^{31}K_{c}$).
      Here we determine the NMR shifts from the peak position of the spectrum.  
    With decreasing temperature, both  NMR shifts increase and show kinks at $T_{\rm N}$.   
      In order to check the magnetic field dependence of the NMR shift in the paramagnetic state, we have measured  $^{31}K_{ab}$ and $^{31}K_{c}$ around 1 T using a resonance frequency of $f$~=~17.235~MHz.  
       No obvious magnetic field dependence of the NMR shift in the paramagnetic state has been observed within our experimental uncertainty,  as shown in Fig.~\ref{fig:Fig2}(c). 
      Similar to the case of $^{59}$Co NMR shift, from $K$-$\chi$ plot analysis,  the hyperfine coupling constants at the P site surrounded by four Co ions are estimated to be $A_{ab}^{\rm P}$ = (5.33 $\pm$ 0.30)~kOe/$\mu_{\rm B}$/Co and $A_{c}^{\rm P}$ = (3.53 $\pm$ 0.17)~kOe/$\mu_{\rm B}$/Co for $H \parallel ab$ and $H \parallel c$, respectively (see, Fig. ~\ref{fig:Fig2}(d)). 
    These values are consistent with the previous report \cite{Imai2017}.
     The orbital shifts are estimated to be $^{31}K_{0,ab}$ = (0.41~$\pm$~0.06)~$\%$ and $^{31}K_{0,c}$ = (0.18~$\pm$~0.02) $\%$, respectively.

\subsection{C.  Magnetic fluctuations in the paramagnetic state}

\begin{figure}[tb]
	\centering
	\includegraphics[width=8.5cm]{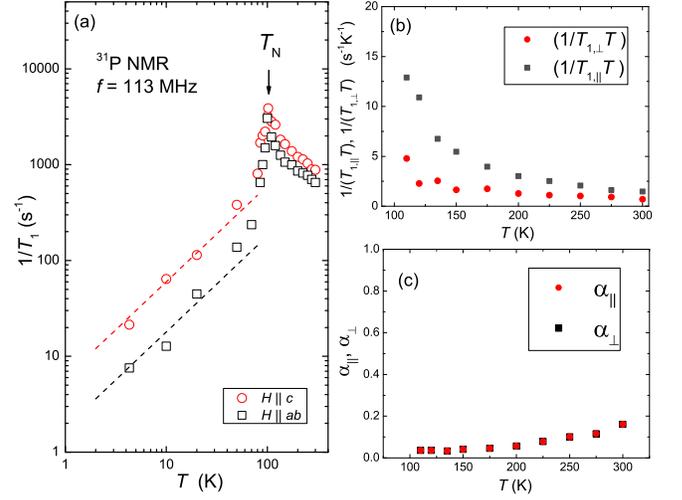} 
	\caption{(Color online)  $T$ dependence of 1/$T_1$ for both magnetic field $H\parallel c$ (open circles) and $H \parallel ab$ (open squares).
The dotted lines represent $1/T_1$ $\propto$ $T$ behavior observed in the AFM state. 
          (b)~$T$ dependence of 1/($T_{1,\perp}T$)  for fluctuations in the $ab$ plane (black) and 1/($T_{1,\parallel}T$)  along the $c$ axis.
		(c)~$T$ dependence of the parameter $\alpha_\perp$ for spin correlations in the $ab$-plane (black) and $\alpha_\parallel$ along the $c$ axis.}
\label{fig:Fig3}
\end{figure}

      In order to investigate the magnetic fluctuations in CaCo$_2$P$_2$, we have measured the $^{31}$P spin-lattice relaxation rate (1/$T_1$) at the peak position of the spectra for both the magnetic field directions. 
     Figure~\ref{fig:Fig3} shows the temperature dependence of 1/$T_1$ where 1/$T_1$ increases with decreasing temperature from 300~K to $T_{\rm N}$ and exhibits a clear peak corresponding to the AFM ordering at $T_{\rm N}$ = 110 K. 
      In the AFM state, 1/$T_1$  for both the magnetic field directions is proportional to $T$, consistent with the metallic state shown by  electrical resistivity measurements \cite{Teruya_LaCo2P2}.

Based on the $T_1$ and the spin part of the Knight shift ($K_{\rm s}$) data,  we discuss the magnetic fluctuations in the paramagnetic state. 
         First we tentatively employ the modified Korringa ratio analysis,  similar to $A$Co$_2$As$_2$ ($A$ =Sr, Ba)~\cite{PaulPRB} and iron pnictide superconductors \cite{PaulPRL}.  
      In Fermi liquid picture, 1/$T_1T$ and $K_{\rm s}$ are determined by the density of state at the Fermi energy $\mathcal{D}(E_{\rm F})$. 
    The $T_1$ has relation with $K_{\rm s}$ that can be described as $T_1TK^2_{\rm s}$ = ($\hbar/4\pi k_{\rm B}$)($\gamma_{\rm e}/\gamma_{\rm n})^2$ $\equiv$ $S$.
       Here $\gamma_{\rm e}$ is electronic gyromagnetic ratio.
      The Korringa ratio $\alpha$ ($\equiv$ $S/T_1TK_{\rm s}$) between an experimental value of $T_1TK^2_{\rm s}$ and the non-interacting electron system $S$ can reveal information about electron correlations in materials \cite{Moriya1963,Narath1968}.
       For uncorrelated electrons, we have $\alpha$$\sim$1.
      However, enhancement of $\chi$ (${\bm q} \neq$ 0) increases 1/$T_1T$ but has little or no effect on $K_{\rm s}$, which probes only the uniform $\chi$ (${\bm q}$ = 0). 
       Thus $\alpha$ $>$ 1 for AFM correlations and $\alpha$ $<$ 1 for ferromagnetic (FM) correlations.
          Since 1/$T_1T$ probes magnetic fluctuations perpendicular to the magnetic field \cite{Moriya1963_2},  it is natural to consider the Korringa ratio 1/($T_{1, \perp}TK^2_{{\rm s}, ab}$), where 1/($T_{1,\perp}T$) = 1/$(T_1T)_{H||c}$, when examining the character of magnetic fluctuations in the $ab$ plane. 
     We also consider the Korringa ratio 1/($T_{1, ||}TK^2_{{\rm s}, c}$) for magnetic fluctuations along the $c$ axis. 
     Here, 1/($T_{1, \parallel}T$) is estimated from 2/$(T_{1}T)_{H||ab}$~$-$~$1/(T_{1}T)_{H||c}$. 
     The temperature dependence of 1/($T_{1, \perp}T$) and 1/$(T_{1, \parallel}T$) above $T_{\rm N}$ are shown in  Fig.~\ref{fig:Fig3}(b).  
     The calculated $\alpha_\parallel$ and $\alpha_\perp$ are shown in Fig.~\ref{fig:Fig3}(c).
     $\alpha_\parallel$ and $\alpha_\perp$ decrease from $\sim$0.2 at 300 K to less than 0.05 around 110 K, indicating dominant ferromagnetic spin correlations between Co spins in the compound.
    Furthermore, no clear difference between $\alpha_\parallel$ and $\alpha_\perp$ suggest that the ferromagnetic fluctuations are nearly isotropic.
       The lowest values of  $\alpha_\parallel$ and $\alpha_\perp$  are almost comparable, or slightly smaller than those in BaCo$_2$As$_2$ and SrCo$_2$As$_2$ in which ferromagnetic fluctuations have been reported \cite{BaCo2As2, PaulPRB}.

\begin{figure}[tb]
	\centering
	\includegraphics[width=8.5cm]{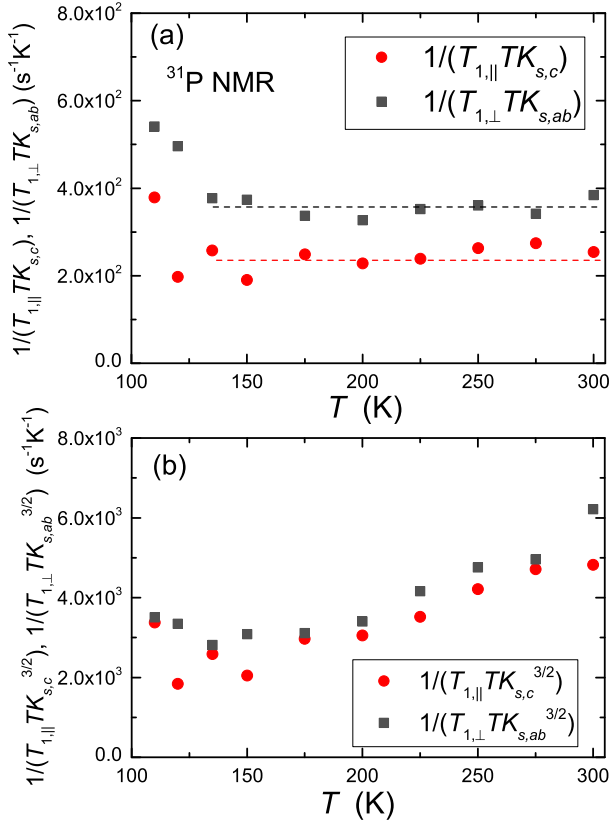} 
	\caption{(Color online) (a)~$T$ dependence of 1/($T_{1,\parallel}TK_{{\rm s},c}$) (red) and 1/($T_{1,\perp}TK_{{\rm s},ab}$) (black). The broken lines are guides for eyes. 
    (b)~$T$ dependence of 1/($T_{1,\parallel}TK_{{\rm s},c}^{3/2}$) (red) and 1/($T_{1,\perp}TK_{{\rm s},ab}^{3/2}$) (black).}
	\label{fig:Fig4}
\end{figure}

    It should be noted that, however, the Korringa analysis usually applies for paramagnetic materials where electron-electron interaction is weak.  
    Since  CaCo$_2$P$_2$ exhibits  an AFM order in contrast to SrCo$_2$As$_2$ and BaCo$_2$As$_2$,  we also analyze NMR data based on self-consistent renormalization (SCR) theory, as Imai ${\it et~ al.}$ have performed \cite{Imai2017}.
    As shown above, the magnetic fluctuations are governed by ferromagnetic spin correlations. 
In this case, according to SCR theory for weak itinerant ferromagnet,  1/$T_1T$ is proportional to $K_{\rm s}$ or to $K^{3/2}_{\rm s}$ for three dimensional (3D) or  two-dimensional (2D) ferrromagnetic spin fluctuations, respectively \cite{SCR1, SCR2}. 
       Figures~\ref{fig:Fig4}(a) and 4(b) show  the $T$ dependence of 1/($T_1TK_{\rm s})$ and 1/($T_1TK_{\rm s}^{3/2})$ for the two directions, respectively.
      Both the 1/($T_{1,\parallel}TK_{\rm s,c}$) and 1/($T_{1,\perp}TK_{{\rm s},ab}$)  are nearly constant in the temperature region of 150 - 300 K and show increases below $\sim$ 150 K, while both the 1/($T_{1,\parallel}TK_{{\rm s},c}^{3/2}$) and 1/($T_{1,\perp}TK_{{\rm s},ab}^{3/2}$) increase with increasing temperature above 150 K. 
      This clearly indicates that  the ferromagnetic spin fluctuations are characterized by 3D  in nature.
     The deviation from the 3D ferromagnetic behavior below 150 K may suggest a change in dimensionality of the ferromagnetic spin correlations from 3D to 2D as can be inferred  from the nearly constant behavior below $\sim$ 150 K shown in  Fig~\ref{fig:Fig4}(b), or is due to an additional contribution from AFM fluctuations close to the A-type AFM ordering  temperature $T_{\rm N}$ = 110 K.


\begin{figure}[tb]
	\centering
	\includegraphics[width=8.5cm]{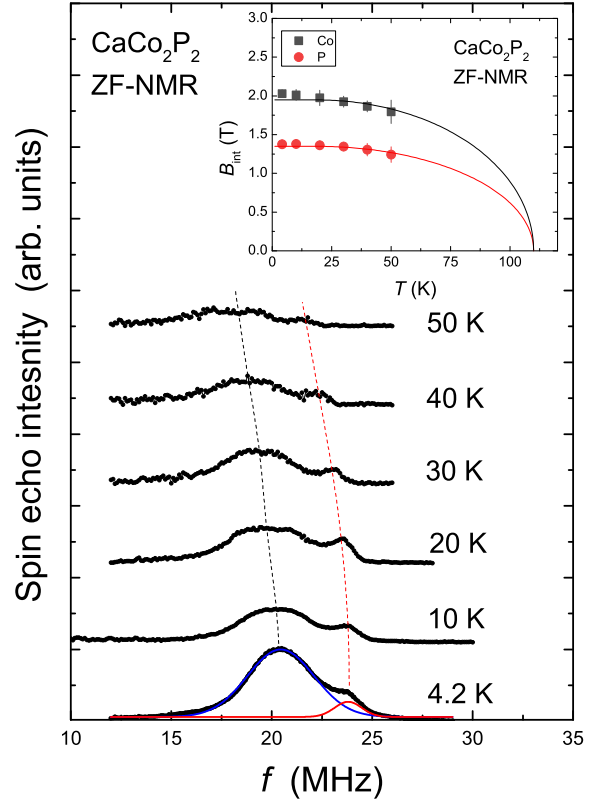} 
	\caption{(Color online) $T$ dependence of ZF-NMR spectrum. At 4.2~K, the blue and red curves represent the $^{59}$Co ZF-NMR  and $^{31}$P ZF-NMR spectra, respectively. The inset shows the $T$ dependence of the internal magnetic induction  $|B_{\rm int}|$ of the $^{59}$Co site (black) and the $^{31}$P site (red). }
	\label{fig:Fig5}
\end{figure}

 \subsection{D.  $^{59}$Co and $^{31}$P zero-field NMR in the antiferromagnetic state}

     Figure \ref{fig:Fig5}  shows $^{59}$Co and $^{31}$P NMR spectra in the AFM ordered state under zero magnetic field at several temperatures up to 50 K. 
     We could not measure the spectra at higher temperatures close to $T_{\rm N}$  due to the reduction of the signal intensity probably originates from shortenings of $T_2$.
     The spectra consist of two broad peaks around 20 MHz and 24 MHz at 4.2 K. 
     Although the peak around 24 MHz is weak, it is  clearly distinguishable and can be observed more clearly at  higher temperatures such as 10 K - 30 K as shown in the figure.
     The broad peak around 20 MHz  is assigned to $^{59}$Co zero-field NMR (ZF-NMR) which does not show any clear quadrupolare splitting due to the inhomogeneous magnetic broadening  as has been observed in the paramagnetic state.
    The weak signal around 24 MHz is attributed to $^{31}$P ZF-NMR, as discussed below. 
    The large difference in the signal intensities  for the two signals is due to the measurement condition which optimizes the signal intensity for $^{59}$Co ZF-NMR. 
    Since the ZF-NMR spectrum originates from the Co ordered magnetic moments, the observation of the ZF-NMR signals is again direct evidence of the magnetic ordered state below  $T_{\rm N}$.

     From the spectrum, the internal magnetic induction $|$$B_{\rm int}^{\rm Co}$$|$ for $^{59}$Co is estimated to be 2.0 T at 4.2 K. 
    $B_{\rm int}^{\rm Co}$ is proportional to $A_{\rm hf}$$<$$\mu$$>$ where $A_{\rm hf}$ is the  hyperfine coupling constant and $<$$\mu$$>$ is the ordered Co magnetic moment. 
      Using $B_{\rm int}^{\rm Co}$ = -2.0 T where the negative sign is reasonably assumed and  $A_{ab}^{\rm Co}$ = (-57.6 $\pm$ 4.2)~kOe/$\mu_{\rm B}$/Co, $<$$\mu$$>$ is estimated to be 0.35 $\mu_{\rm B}$ which is in good agreement with 0.32 $\mu_{\rm B}$ reported by the ND measurement \cite {Reehuis1998} and slightly smaller than 0.4$\mu_{\rm B}$ from $\mu$SR measurements \cite{Sugiyama2015PRB}. 
      Here we used the hyperfine coupling constant for the $ab$ plane direction since the Co moments are in the $ab$ plane as will be shown below.

      In the case of $^{31}$P ZF-NMR, the internal magnetic induction $|$$B_{\rm int}^{\rm P}$$|$ for $^{31}$P is estimated to be 1.4 T at 4.2 K. 
      From  $A_{ab}^{P}$ = (5.33$\pm$ 0.30)~kOe/$\mu_{\rm B}$/Co, $<$$\mu$$>$ is estimated to be 0.65 $\mu_{\rm B}$ which is much greater than the reported values and also  the estimated value from the Co NMR data.
     At present, the reason for the estimated large value from $^{31}$P NMR results is not clear, but it may suggest that  the $A^{\rm P}_{ab}$ in the AFM state is slightly greater than that  in the PM state.
    This would be possible if one considers the effects of the next-nearest-neighbor (NNN) Co spins which are assumed to produce a negative hyperfine field at the P site. 
     Since in the AFM state  the direction of NNN Co spins is antiparallel to that of the NN Co spins, one expects a positive hyperfine field at the P sites from the NNN Co spins,  which increases the positive hyperfine field produced by the NN Co spins.  
     Assuming  $<$$\mu$$>$ = 0.32~$\mu_{\rm B}$ (Ref. \onlinecite{Reehuis1998}), we thus estimate a $\sim$ 25\% additional contribution of the hyperfine field from the NNN Co spins to the total hyperfine field at the P site.

     The temperature dependence of  $|B_{\rm int}^{\rm Co}|$ and  $|B_{\rm int}^{\rm P}|$ is shown in the inset of Fig. \ref{fig:Fig5} where  both show smooth decrease with increasing temperature without any anomalies around $T^*$ $\sim$ $32-36$ K where the magnetic susceptibility shows a maximum.   
     This indicates that the reduction of $\chi$ below $T^*$ is not due to a change in the magnitude of the Co ordered moments at $T^*$. 
     The temperature dependence of $|B_{\rm int}^{\rm Co}|$ and  $|B_{\rm int}^{\rm P}|$ is well reproduced by a Brillouin function which was calculated based on the Weiss molecular field model with $S$ = 3/2 for Co$^{2+}$, $T_{\rm N}$ = 110 K, $|B_{\rm int}^{\rm Co}|$  = 2.0~T and $B_{\rm int}^{\rm P}$ = 1.4 T at $T$ = 4.2~K [solid curves in the inset of Fig.~\ref{fig:Fig5}]. 
   These results indicate that  the magnetic state of the Co ions is well explained by the local moment picture although the system is metallic as determined from electrical resistivity measurements \cite{Teruya_LaCo2P2} and also from 1/$T_1T$ = const. behavior in the AFM state.

\begin{figure}[tb]
	\centering
	\includegraphics[width=8.0cm]{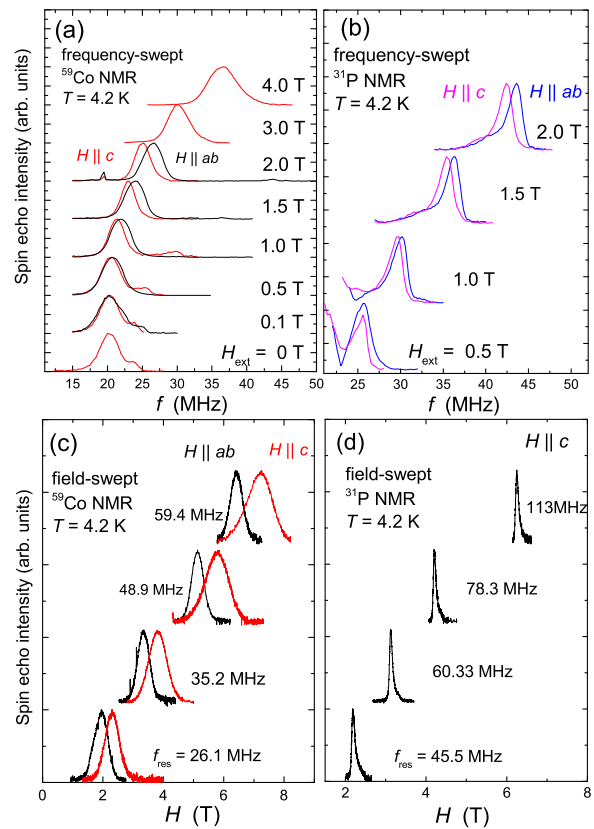} 
	\caption{(Color online) (a) Frequency-swept $^{59}$Co-NMR spectra at $T$ = 4.2 K under several magnetic fields parallel to the $c$ axis (red) and to the $ab$ plane  (black). (b) Similar frequency-swept $^{31}$P-NMR spectra under magnetic fields at $T$ = 4.2 K [$H \parallel c$  axis (pink) and $H \parallel ab$ plane  (blue)].
 It is noted that the signals below $\sim$ 23 MHz for $H_{\rm ext}$ = 0.5 T originate from $^{59}$Co NMR. 
       (c) Field-swept $^{59}$Co-NMR spectra for $H \parallel c$  axis (red) and $H \parallel ab$ plane  (black). (d) Field-swept $^{31}$P-NMR spectra for $H\parallel c$ axis.
}
	\label{fig:Fig6}
\end{figure}

\begin{figure}[tb]
	\centering
	\includegraphics[width=8.5cm]{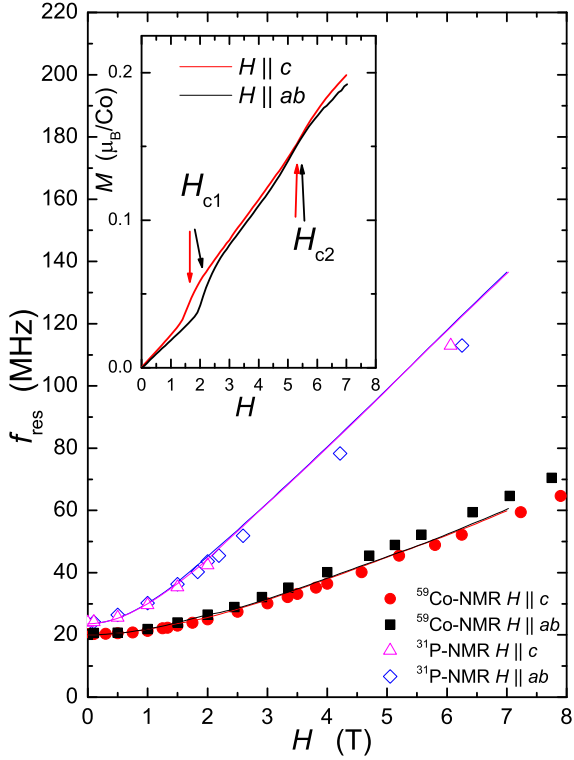} 
	\caption{(Color online)  $H$ dependence of resonance frequency ($f_{\rm res}$) for both nuclei $^{59}$Co (closed symbols) and $^{31}$P (open symbols).   $f_{\rm res}$s are mainly determined  from the frequency-swept NMR spectra and the $H$-swept NMR spectra below and above $\sim$ 3 T, respectively. 
    The black (red) and the blue (pink) curves are calculated results for $^{59}$Co and $^{31}$P NMR for $H\parallel ab$ ($H\parallel c$), respectively. 
    The two curves for each nucleus are nearly same. 
		Inset: magnetization curves at 2~K for both magnetic field directions from Ref. \onlinecite{Teruya_LaCo2P2}.
Note that metamagnetic-like behaviors are observed at $H_{\rm c1}$ $\sim$ 2.1 T (1.6 T) and $H_{\rm c2}$ $\sim$ 5.1 T (5.1 T) for $H \parallel c$ ($H \parallel ab$).
}
	\label{fig:Fig6b}
\end{figure}

 \subsection{E. $^{59}$Co and $^{31}$P NMR in the antiferromagnetic state}

     Now we show the external field dependence of $^{59}$Co and $^{31}$P NMR spectra  in the AFM state at 4.2 K.
     Figures \ref{fig:Fig6}(a) and  \ref{fig:Fig6}(b) show frequency-swept $^{59}$Co NMR and  $^{31}$P NMR spectrum, respectively, measured under several external magnetic fields with the two different directions $H \parallel c$ and $H \parallel  ab$.
      In AFM state, one expects a splitting of NMR line when external magnetic field  is applied along magnetic easy axis, while only shifting of NMR line without splitting is expected when $H$ is applied to perpendicular to the magnetic easy axis, for $H$ smaller than magnetocrystalline anisotropy field. 
      On the other hand, when $H$ is greater than magnetocrystalline anisotropy field, magnetic moments change the direction to perpendicular to $H$, known as spin flop.  
     As shown in Fig. \ref{fig:Fig6}(a), no splitting of the $^{59}$Co NMR line is observed less than $\sim$ 0.1 T for the two magnetic field directions.
     This indicates that the magnetic anisotropy field of the Co magnetic moments is less 0.1 T and the application of magnetic field makes a spin flop very easily.
     In fact, as shown in the inset of Fig. \ref{fig:Fig6b}, the magnetization vs. external magnetic field plot  exhibits very similar behavior for both the magnetic field directions, suggesting the very weak magnetic anisotropy,   although metamagnetic-like behaviors are observed around 2.1 T and 5.1 T for $H \parallel c$ and around 1.6 T and 5.1 T for $H$$\perp$$c$.

\begin{figure}[tb]
	\centering
	\includegraphics[width=8.5cm]{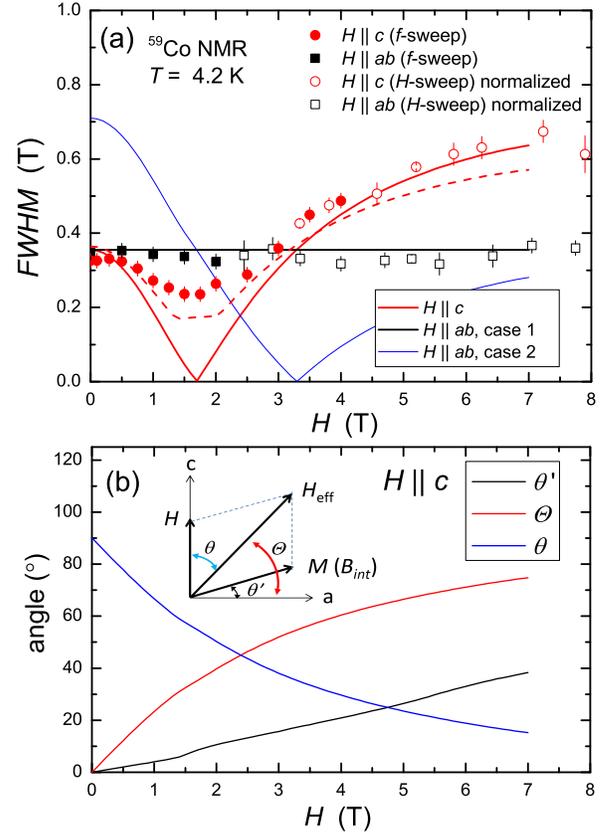} 
	\caption{(Color online)  (a) External magnetic field $H$ dependence of the full-width at half maximum ($FWHM$) of the field-swept $^{59}$Co-NMR spectrum for magnetic fields $H\parallel c$ axis (red) and $H \parallel ab$ plane (black).
        The curves are calculated results (see text).   
    (b) $H$ dependence of angles $\theta^{\prime}$, $\Theta$ and $\theta$ for the case of $H$ $\parallel$ $ c$ axis.
Here we used the magnetization data measured at $T$ = 2 K. 
	Inset: a schematic  view of  the angles $\theta^{\prime}$, $\Theta$ , and $\theta$. Note that the direction of $B_{\rm int}$ for Co is antiparallel to that of the magnetization $M$ while that for P is parallel to the $M$. }
	\label{fig:Fig7}
\end{figure}

     Figure \ref{fig:Fig6b} shows the external field dependence of resonance frequencies  ($f_{\rm res}$) for frequency-swept  $^{59}$Co and $^{31}$P NMR spectra determined from the peak position of each spectrum.
     We also measured  magnetic field-swept  $^{59}$Co and $^{31}$P NMR spectra at several constant resonance frequencies above around 2 T [see, Figs. \ref{fig:Fig6}(c) and (d)] whose results are also plotted in the figure.  
         The $f_{\rm res}$s are nearly constant below 1 T and increase gradually with increasing $H$. 
           The resonance frequency is proportional to an effective field ($H_{\rm eff}$) which is the vector sum of the internal magnetic  induction $\bf{B}_{\rm int}^{\rm Co}$ and the external field $\bf{H}$,  i.e., $|$$\bf{H}_{\rm eff}$$|$ = $|$$\bf{B}_{\rm int}^{\rm Co}$ + $\bf{H}$$|$. 
Therefore,  the resonance frequency $f_{\rm res}$ is expressed  as  
\begin{eqnarray}
\centering
f_{\rm res}  = \frac{\gamma_{\rm n}}{2\pi}  H_{\rm eff} = \frac{\gamma_{\rm n}}{2\pi}  \sqrt{\mathstrut H^2 + B_{\rm int}^2 + 2HB_{\rm int}{\rm sin}\theta^{\prime} }.
\label{eq:3}
\end{eqnarray} 
Here $\theta^{\prime}$ is the canting angle of the Co ordered moment from the perpendicular direction  with respect to the external magnetic field [see, Fig \ref{fig:Fig7}(b)], which can be expressed as $\theta^{\prime}$ = sin$^{-1}$($M/M_{\rm s})$ where $M_{\rm s}$ is the saturation of magnetization.

     From the magnetization curves from Ref. \onlinecite{Teruya_LaCo2P2} shown in the inset of Fig. \ref{fig:Fig6b}, we calculated the external field dependence of sin$\theta^{\prime}$ = $M/M_{\rm s}$ for $H \parallel c$ and $H \parallel  ab$ where we used $M_{\rm s}$ = 0.32 $\mu_{\rm B}$ \cite{Reehuis1998}. 
     Using  $B_{\rm int}^{\rm Co}$ = $-$2.0 T, one can calculate the external field dependence of the resonance frequency. 
   The black and red solid lines representing the calculated $H$ dependence of $f_{\rm res}$  for $^{59}$Co NMR   under the two magnetic field directions reproduce experimental results relatively well. 
    Here we assumed the internal magnetic induction is isotropic for simplicity since the $^{59}$Co hyperfine coupling constant estimated above is nearly isotropic.
    No clear anomalies due to the metamagnetic-like  behavior around 2 T  are observed in the external field dependence of the resonance frequencies for not only experimental results but also in the calculated one.
     Assuming the metamagnetic-like behavior originates from a change in the direction of the magnetization (or the Co ordered moments), the changes are estimated to be $\sim$5$^{\circ}$ along the magnetic field direction  from the magnetization data, which cannot be detected within our experimental uncertainty.
     Similar to the $^{59}$Co NMR, the external field dependence of resonance frequency for $^{31}$P NMR can be reasonably reproduced by Eq. (3) using  $B_{\rm int}^{\rm P}$ = 1.4 T obtained from $^{31}$P ZF-NMR as shown by the blue and pink curves.

     Figure \ref{fig:Fig7}(a) shows the $H$ dependence of $FWHM$ of $^{59}$Co NMR spectra for $H \parallel c$ and $H \parallel ab$.
     The filled and open  symbols represent the results from the frequency-swept and field-swept $^{59}$Co-NMR spectra, respectively.
     In the case of the field-swept NMR spectrum, the plotted values are reduced by a factor of 1.4 to match the results from the frequency-swept NMR measurement. 
     This is due to the fact that the field-swept NMR spectrum measurements give rise to greater $FWHM$ than those of the frequency-swept NMR spectra because the slope of $f_{\rm res}$ vs. $H$ is less than the value of $\gamma_{\rm n}$/2$\pi$.
    It is noted that one cannot estimate the $FWHM$ of the $^{59}$Co-NMR spectrum at low magnetic fields less than $\sim$ 1 T by sweeping magnetic field. 
     Although the values of $FWHM$ from the field-swept NMR spectra are slightly greater, the intrinsic behavior of the $H$ dependence of $FWHM$ is not affected, as actually be seen in the $H \parallel c$ data in the magnetic field region of $H$ = $ 2 - 4 $ T where both data sets exhibit the same magnetic field dependence.

    As shown in Fig. \ref{fig:Fig7}(a),  $FWHM$ $\sim$ 0.36 T is  nearly independent of $H$ for $H \parallel ab$.
    On the other hand, in the case of $H \parallel c$, with increasing $H$  the $FWHM$ decreases gradually and show a local minimum around  1.6 T and then starts to increase.   
   No change in the $FWHM$ for $H \parallel ab$ indicates that the degree of the inhomogeneous magnetic broadening is independent of $H$.   
   Therefore, it is clear that the characteristic behavior of $FWHM$ for $H \parallel c$ is not due to the magnetic broadening effect. 
   Here we attribute it to the effects of quadrupole splitting of the spectrum.
    As described in section III.A,  the $FWHM$ depends on $\theta$ as $FWHM \propto$ $|$3cos$^2\theta$-1$|$ where $\theta$ is the angle between the principal axis of the EFG and the quantization axis of nuclear spin ($H_{\rm eff}$). 
      When $H$ is applied along the $c$ axis (i.e. parallel to the principal axis of the EFG) in the AFM state,  the angle $\theta$ changes  by the application of magnetic field [see, Fig. \ref{fig:Fig7}(b)].
     When $H$ is very small compared with $B_{\rm int}$, $\theta$ is close to 90$^{\circ}$. 
     With increasing  $H$,  $\theta$ decreases and becomes close to 54.7 $^\circ$ where  $FWHM$ is expected to be a minimum.
     With further increase of  $H$,  $\theta$ will be decreased  and eventually becomes close to zero, resulting in spectrum twice broader compared with that for $\theta$ = 90$^{\circ}$.
     This scenario, the change in $\theta$ with $H$, qualitatively  explains the $H$ dependence of $FWHM$.

    To analyze the experimental results more quantitatively, we have calculated the $H$ dependence of $FWHM$.
     In  the case of $H \parallel c$, utilizing $\theta^{\prime}$ estimated from the magnetization data, one can easily calculate  the $H$ dependence of  $\Theta$ which is the angle between the direction of $H_{\rm eff}$ and  the $ab$ plane, and thus also the angle $\theta$ between the $c$ axis and $H_{\rm eff}$ corresponding to the quantization axis for nuclear spin.
    The calculated $H$ dependence of $\Theta$, $\theta$ and $\theta^{\prime}$ for $H \parallel c$ is shown in Fig. \ref{fig:Fig7}(b) for $T$ = 2 K.
    We simply assumed that the $FWHM$ can be written as  $FWHM$ = $a$$|$3cos$^2\theta-1|$. 
    The solid red curve in Fig. \ref{fig:Fig7}(a) shows the calculated result with $a$ = 0.36 T, which reproduces the experimental data well, although one can see the deviation around 1.5 T. 
   The deviation can be due to inhomogeneous magnetic broadening of each line and also a misalignment between the $H$ and the $c$ axis since the simple model does not take such effects into consideration.
   In fact, if we take the magnetic broadening of each line (0.1 T) and the misalignment of 5 $^{\circ}$, one can reproduce the results a little bit better as shown by the red dashed line.
    In the case of $H \parallel ab$, the nearly constant $FWHM$ $\sim$ 0.36 T can be also well reproduced with $a$ = 0.36 T and $\theta$ = 90$^{\circ}$, as shown by the solid black line (case 1) in the figure.
     This indicates that $H_{\rm eff}$ is always in the $ab$ plane keeping $\theta$ = 90$^{\circ}$ and the spin-flip occurs in the $ab$ plane. 
      If one assumes that the Co ordered moments flip to the $c$-axis direction when $H$ is applied in the $ab$ plane, the $FWHM$ should depend on  $H$  as shown by the solid blue curve (case 2)  which clearly contradicts with the experimental results.
       Thus, these results indicate the $ab$ plane is the magnetic easy plane. 
       As the $ab$-plane magnetization starts to increase from nearly zero $H$, similar to the case of the $c$-axis magnetization as shown in the inset of Fig. \ref{fig:Fig6b}, the in-plane magnetic anisotropy is considered to be also very weak.
    To study the details of the characteristic properties of the magnetic anisotropy in CaCo$_2$P$_2$, it is important and highly required to measure the magnetization for both magnetic field directions in detail at low magnetic fields below 0.1 T. 
    This is the future project.
      The $FWHM$ ($\sim$  0.36 T) of the $^{59}$Co ZF-NMR corresponds to the case for $\theta$ = 90$^{\circ}$. 
    This evidences that the Co ordered moments are in the $ab$ plane at zero magnetic field, again consistent with  the ND results  \cite{Reehuis1998}.

\begin{figure}[tb]
	\centering
	\includegraphics[width=8.5cm]{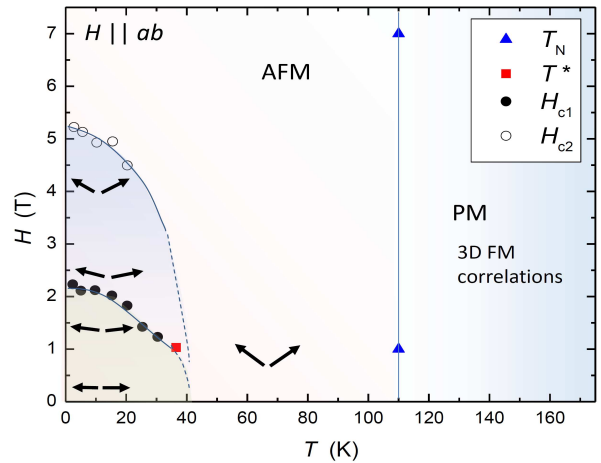} 
	\caption{(Color online)  (a) External magnetic fied $H$ - temperature magnetic phase diagram for $H \parallel ab$ plane.
$H_{c1}$,  $H_{c2}$ and $T^*$ are from Ref. \onlinecite{Teruya_LaCo2P2}. }
	\label{fig:Fig8}
\end{figure}

 \subsection{F.  Magnetic phase diagram}

    Figure 9 summarizes the magnetic phase diagram of CaCo$_2$P$_2$ for $H \parallel ab$, based on the present $^{59}$Co and $^{31}$P NMR studies  and also the previous magnetization measurements \cite{Teruya_LaCo2P2}.
     Under zero magnetic field, the system is the A-type AFM state below $T_{\rm N}$ = 110 K where the Co ordered moment are in the $ab$ plane.
  
     When $H$ is applied in the $ab$ plane, the Co ordered moments flop immediately to make the perpendicular configuration between it and $H$,  and then cant along the $H$ direction  with increasing $H$, producing the linear increase of the magnetization with increasing $H$ as actually observed (see, Fig. \ref{fig:Fig6}).    
      When  $H$ reaches $H_{c1}$, the canting angle $\theta^{\prime}$ jumps by  $\sim$5 degree at $H_{c1}$ at 2 K, producing the metamagnetic behavior in the magnetization curve.  
     With further increasing $H$, the $\theta^{\prime}$ keeps increasing with a small jump at $H_{c2}$ with  a change of the canting angle less than $\sim$4 degree at 2 K.  

   In the case of $H \parallel c$,  with increasing $H$ the canting angle $\theta^{\prime}$ increases, similar to the case of $H \parallel ab$ although no spin flop occurs because of the perpendicular configuration between the Co ordered moments and $H$ at the initial condition.
    With further increase of $H$, the canting angle jumps slightly at $H_{c1}$ and $H_{c2}$ similar to the case of $H$ $\parallel$ $ab$, producing again the  metamagnetic behavior in the magnetization curve.
    The changes in $\theta^{\prime}$ are close to $\sim$4 degree for each  jump.           
    The similar slopes in the magnetization curves for $H$ parallel and perpendicular to the $c$ axis indicate that the in-plane and out-of-plane magnetic anisotropies are nearly the same.  

     Finally we would like to comment on the reduction of the magnetic susceptibility below $T^*$. 
     It is evidenced that, from the temperature dependence of $B_{\rm int}$,  there is no significant change in the magnitude of the Co ordered moment below $T^*$.
    Therefore, the reduction of the magnetic susceptibility below $T^*$ cannot be due to a reduction of the Co ordered moments.  
    In addition, we do not observe any anomaly around $T^*$ in the temperature dependence of $^{59}$Co $T_1$ under zero magnetic field (not shown), suggesting no obvious phase transition at $T^*$.
    Our interpretation is that the canting angle decreases below $T^*$. 
    This corresponds to a reduction of the relative angle of the ferromagnetically ordered Co moments between the adjacent  Co layers,
   making  a decrease in the magnetic susceptibility as has been observed at $T^*$ in CaCo$_2$P$_2$.

 \section{IV.  Summary}
     We have reported a comprehensive  study of $^{59}$Co and $^{31}$P NMR measurements in external magnetic field and zero magnetic field in the paramagnetic and the AFM states in the A-type AFM  CaCo$_2$P$_2$ with a N\'eel temperature of $T_{\rm N}\sim$110~K.
    The AFM transition has been confirmed from NMR data, especially, the relaxation rates 1/$T_1$ exhibiting a clear peak at at $T_{\rm N}$.
    The magnetic fluctuations in the paramagnetic state was found to be three-dimensional ferromagnetic from the analysis of NMR data, suggesting ferromagnetic interaction between Co spins in the ${\it ab}$ plane characterize the spin correlations in the paramagnetic state.   
     In the AFM state below $T_{\rm N}$, we have observed $^{59}$Co and $^{31}$P NMR signals under zero magnetic field.
     The analysis of $^{59}$Co ZF-NMR spectrum and also  the external field dependence of $FWHM$ of $^{59}$Co NMR spectrum,  the Co ordered moments are found to be in the $ab$ plane and are estimated to be 0.35 $\mu_{\rm B}$ at  4.2  K. 
    The behaviors of spin canting of the Co ordered moments produced by  the application of external magnetic field have been clearly observed from a microscopic point of view based on the $H$ dependence of $FWHM$.    
     Furthermore, the external field dependence of $^{59}$Co NMR spectrum in the AFM state suggests a very weak  magnetic  anisotropy of the Co ions.  
     The magnetic state of the Co ions in CaCo$_2$P$_2$  is well explained by the local moment picture in the AFM state without showing any anomalies at $T^*$ = 32-36 K  where magnetic susceptibility exhibits a maximum. 
     This indicates that the magnitude of the Co ordered moments does not change at $T^*$. 
     We attributed the reduction of $\chi$ below $T^*$ to the decrease in the canting angle of the Co ordered moments.
     The scenario also reasonably explains the metamagnetic-like behavior observed in the magnetization curves.  
      At present, however, the origin of the change in the canting angle at $T^*$ and also $H_{c1}$ and $H_{c2}$ is not clear. 
     Further studies, especially theoretical works, are highly required  to shed light on the peculiar magnetic properties of CaCo$_2$P$_2$.

 \section{V.  Acknowledgments} 
    We thank H. Uehara and F. Kubota for assistance with the experiments.
    The research was supported by the U.S. Department of Energy, Office of Basic Energy Sciences, Division of Materials Sciences and Engineering. Ames Laboratory is operated for the U.S. Department of Energy by Iowa State University under Contract No.~DE-AC02-07CH11358.
 Part of the work was supported by the Japan Society for the Promotion of Science KAKENHI: J-Physics (Grant Nos. JP15K21732, JP15H05885, and JP16H01078).
 N. H. also thanks the KAKENHI: J-Physics for financial support to be a visiting scholar at the Ames laboratory.


\begin{thebibliography}{99}


\bibitem{Kamihara2008} Y. Kamihara, T. Watanabe, M. Hirano, and H. Hosono, J. Am. Chem. Soc. {\bf 130}, 3296 (2008).
\bibitem{Johnston2010} D. C. Johnston, Adv.  Phys. {\bf 59}, 803 (2010).
\bibitem{Canfield2010} P. C. Canfield and S.~ L. Bud'ko, Annu. Rev. Condens. Matter Phys. {\bf 1}, 27 (2010).
\bibitem{Stewart2011} G.  R. Stewart, Rev. Mod. Phys. {\bf 83}, 1589 (2011).
\bibitem{Nakai2008} Y. Nakai, K. Ishida, Y. Kamihara, M. Hirano, and H. Hosono, Phys. Rev. Lett. {\bf 101}, 077006 (2008).
\bibitem{PaulPRB} P. Wiecki, V. Ogloblichev, A. Pandey, D. C. Johnston,  and Y. Furukawa,  Phys. Rev. B {\bf 91}, 220406(R) (2015).
\bibitem{PaulPRL} P. Wiecki, B. Roy, D. C. Johnston, S. L. Bud'ko, P. C. Canfield, and Y. Furukawa,  Phys. Rev. Lett. {\bf 115}, 137001 (2015).
\bibitem{Cui2017} J. Cui, Q.-P. Ding, W. R. Meier, A. E. B\"ohmer, T. Kong, V. Borisov, Y. Lee, S. L. Bud'ko, R. Valent\'i, P. C. Canfield, and Y. Furukawa, Phys. Rev. B  {\bf96}, 104512 (2017).
\bibitem{Hoyer2016} M. Hoyer, R. M. Fernandes, A. Levchenko, and  J. Schmalian, Phys. Rev. B {\bf 93}, 144414 (2016).
\bibitem{Allred2016} J. M. Allred, K. M. Taddei, D. E. Bugaris, M. J. Krogstad, S. H. Lapidus, D. Y. Chung, H. Claus, M. G. Kanatzidis, D. E. Brown, J. Kang, R. M. Fernandes, I. Eremin, O. Chmaissem, and R. Osborn, Nat. Phys. {\bf 12}, 493 (2016).
\bibitem{Kim2010} M. G. Kim, A. Kreyssig, A. Thaler, D. K. Pratt, W. Tian, J. L. Zarestky, M. A. Green, S. L. Bud'ko, P. C. Canfield, R. J. McQueeney, and A. I. Goldman,  Phys. Rev. B {\bf 82}, 220503(R) (2010).
\bibitem{Avci2014} S. Avci, O. Chmaissem, J. M. Allred, S. Rosenkranz, I. Eremin, A. V. Chubukov, D. E. Bugaris, D. Y. Chung, M. G. Kanatzidis, J.-P. Castellan, J. A. Schlueter, H. Claus, D. D. Khalyavin, P. Manuel, A. Daoud-Aladine, and R. Osborn, Nat. Commun. {\bf 5}, 3845 (2014).
\bibitem{Wang2016} L. Wang, F. Hardy, A. E. B\"ohmer, T. Wolf, P. Schweiss, and C. Meingast, Phys. Rev. B {\bf 93}, 014514 (2016).
\bibitem{Hassinger2012} E. Hassinger, G. Gredat, F. Valade, S. R. de Cotret, A. Juneau-Fecteau, J.-Ph. Reid, H. Kim, M. A. Tanatar, R. Prozorov, B. Shen, H.-H. Wen, N. Doiron-Leyraud, and L. Taillefer,  Phys. Rev. B {\bf 86}, 140502(R) (2012).
\bibitem{Bohmer2015} A. E. B\"ohmer, F. Hardy, L. Wang, T. Wolf, P. Schweiss, and C. Meingast, Nat. Commun. {\bf 6}, 7911 (2015).
\bibitem{Allred2015} J. M. Allred, S. Avci, D. Y. Chung, H. Claus, D. D. Khalyavin, P. Manuel, K. M. Taddei, M. G. Kanatzidis, S. Rosenkranz, R. Osborn, and O. Chmaissem,  Phys. Rev. B {\bf 92}, 094515 (2015).
\bibitem{Hassinger2016} E. Hassinger, G. Gredat, F. Valade, S. R. de Cotret, O. Cyr-Choini\`ere, A. Juneau-Fecteau, J.-Ph. Reid, H. Kim, M. A. Tanatar, R. Prozorov, B. Shen, H.-H. Wen, N. Doiron-Leyraud, and L. Taillefer, Phys. Rev. B {\bf 93}, 144401 (2016).
\bibitem{Meier20172} W. R. Meier, Q.-P. Ding, A. Kreyssig, S. L. Bud'ko, A. Sapkota, K. Kothapalli, V. Borisov, R. Valent\'i, C. D. Batista, P. P. Orth, R. M. Fernandes, A. I. Goldman, Y. Furukawa, A. E. B\"ohmer, and P. C. Canfield,  npj Quant. Mater. {\bf 3,} 5 (2018).
\bibitem{DingPRB2017} Q.-P. Ding, W. R. Meier,  A. E. B\"ohmer, S. L. Bud'ko, P. C. Canfield, and Y. Furukawa,  Phys. Rev. B {\bf 96}, 220510(R) (2017).
\bibitem{Kreyssig2018} A. Kreyssig, J. M. Wilde, A. E.  B\"ohmer, W. Tian, W. R. Meier, B. Li, B. G. Ueland, M. Xu, S. L. Bud'ko, P. C. Canfield, R. J. McQueeney, and A. I. Goldman, Phys. Rev. B {\bf 97}, 224521 (2018).

\bibitem{Kreyssig2008} A. Kreyssig, M. A. Green, Y. B. Lee, G. D. Samolyuk, P. Zajdel, J. W. Lynn, S. L.  Bud'ko, M. S. Torikachvili, N. Ni, S. Nandi, J. B. Le${\rm \tilde{a}}$o, S. J. Poulton, D.  N. Argyriou, B.  N. Harmon, R.  J. McQueeney, P. C. Canfield, and A. I. Goldman, Phys. Rev. B {\bf 78}, 184517 (2008).
\bibitem{Pratt2009}  D. K. Pratt, W. Tian, A. Kreyssig, J. L. Zarestky, S. Nandi, N. Ni, S. L. Bud'ko, P. C. Canfield, A. I. Goldman, and R. J. McQueeney, Phys. Rev. Lett. {\bf 103}, 087001 (2009).
\bibitem{Prokes2010} K. Proke${\rm \breve{s}}$, A. Kreyssig, B. Ouladdiaf, D. K. Pratt, N. Ni, S. L. Bud'ko, P. C. Canfield, R. J. McQueeney, D. N. Argyriou, and A. I. Goldman, Phys. Rev. B {\bf 81}, 180506(R) (2010).
\bibitem{Ran2011} S. Ran, S. L.  Bud'ko, D. K. Pratt, A. Kreyssig, M. G. Kim, M. J. Kramer, D. H. Ryan, W. N. Rowan-Weetaluktuk, Y. Furukawa, B. Roy, A. I. Goldman, and P. C. Canfield, Phys. Rev. B {\bf 83}, 144517 (2011).

\bibitem{Kawasaki2010} S. Kawasaki, T. Tabuchi, X. F. Wang, X. H. Chen, G-q. Zheng, Supercond. Sci. Technol. {\bf 23}, 054004 (2010). 
\bibitem{Soh2013} J. H. Soh, G. S. Tucker, D. K. Pratt, D. L. Abernathy, M. B. Stone, S. Ran, S. L.  Bud'ko, P. C. Canfield, A. Kreyssig, R. J. McQueeney, and A. I. Goldman, Phys. Rev. Lett. {\bf 111}, 227002 (2013).
\bibitem{Furukawa2014} Y. Furukawa, B. Roy, S. Ran, S. L. Bud'ko, and P. C. Canfield, Phys. Rev. B {\bf 89}, 121109 (2014).

\bibitem{Jia2009} S. Jia, A. J. Williams, P. W. Stephens, and R. J. Cava, Phys. Rev. B {\bf 80}, 165107 (2009).
\bibitem{Moresen1998} E. Morsen, B. Mosel, W. Muller-Warmuth, M. Reehuis, and W. Jeitschko, J. Phys. Chem. Solids {\bf 49}, 785 (1988).
\bibitem{Imai2014} M. Imai, C. Michioka, H. Ohta, A. Matsuo, K. Kindo, H. Ueda, and K. Yoshimura, Phys. Rev. B {\bf 90}, 014407 (2014).
\bibitem{Imai2015PP} M. Imai, C. Michioka, H. Ueda, A. Matsuo, K. Kindo, and K. Yoshimura, Phys. Procedia 75, 142 (2015).
\bibitem{Imai2017} M. Imai, C. Michioka, H. Ueda and K. Yoshimura, Phys. Rev. B {\bf 95}, 054417 (2017).

\bibitem{Reehuis1994} M. Reehuis, C. Ritter, R. Ballou, and W. Jeitschko, J. Magn. Magn. Mater. {\bf 138}, 85 (1994).

\bibitem{Imai2015PRB} M. Imai, C. Michioka, H. Ueda, and K. Yoshimura, Phys. Rev. B 91, 184414 (2015).

\bibitem{Teruya_LaCo2P2} A. Teruya, A.Nakamura, T. Takeuchi, F. Honda, D. Aoki, H. Harima, K. Uchima, M. Hedo, T. Nakama and Y. \=Onuki, Phys. Procedia. {\bf 75}, 876 (2015).

\bibitem{Reehuis1998}  M. Reehuis, W. Jeitschko, G. Kotzyba, B. Zimmer, and X. Hu, J. Alloys Compd. {\bf 266}, 54 (1998).

\bibitem{Baumbach2014PRB} R. E. Baumbach, V. A. Sidorov, Xin Lu, N. J. Ghimire, F. Ronning, B. L. Scott, D. J. Williams, E. D. Bauer, and J. D. Thompson, Phys. Rev. B {\bf 89}, 094408 (2014).
\bibitem{Slichter_book} C. P. Slichter, {\it Principles of Magnetic Resonance}, 3rd ed. (Springer, New York, 1990).

\bibitem{YVO3} J. Kikuchi, H. Yasuoka, Y. Kokubo, and Y. Ueda, J. Phys. Soc. Jpn, {\bf 63} 3577 (1994).
\bibitem{Vasily2010}  V. Ogloblichev, K. Kumagai, S. Verkhovskii, A. Yakubovsky, K. Mikhalev, Y. Furukawa, A. Gerashenko, A. Smolnikov, S. Barilo, G. Bychkov, and S. Shiryaev, Phys. Rev. B {\bf 81}, 144404 (2010).

\bibitem{Abragam1955} A. Abragam, J. Horowitz, and M. H. L. Pryce, Proc. Roy. Soc. (London), Ser. A {\bf 230}, 169 (1955).

\bibitem{Tsuda1968} T. Tsuda, A. Hirai, and H. Abe, Phys. Lett. {\bf 26A}, 463 (1968).
\bibitem{Fukai1996} T. Fukai, Y. Furukawa, S. Wada, and K. Miyatani, J. Phys. Soc. Jpn. {\bf 65}, 4067 (1996).
\bibitem{Roy2013} B. Roy, Abhishek Pandey, Q. Zhang, T. W. Heitmann, D. Vaknin,  D. C. Johnston, and Y. Furukawa, Phys. Rev. B {\bf 88}, 174415  (2013). 


\bibitem{Moriya1963} T. Moriya, J. Phys. Soc. Jpn. {\bf 18}, 232 (1963). 
\bibitem{Narath1968}A. Narath and H. T. Weaver, Phys. Rev. {\bf 175}, 378 (1968).
\bibitem{Moriya1963_2} T. Moriya, J. Phys. Soc. Jpn. {\bf 18}, 516 (1963).

\bibitem{BaCo2As2}  K. Ahilan, T. Imai, A. S. Sefat, and F. L. Ning, Phys. Rev. B {\bf 90}, 014520 (2014).

\bibitem{SCR1} M. Hatatani and T. Moriya, J. Phys. Soc. Jpn. {\bf 64}, 3434 (1995).
\bibitem{SCR2} T. Moriya, Spin~Fluctuations~in~Itinerant~Electron~Magnetism, Vol. 56 (Springer-Verlag, 1985).

\bibitem{Sugiyama2015PRB} J. Sugiyama, H. Nozaki, I. Umegaki, M. Harada, Y. Higuchi, K. Miwa, E. J. Ansaldo, J. H. Brewer, M. Imai, and C. Michioka, K. Yoshimura, and M. Maånsson, Phys. Rev. B {\bf 91}, 144423 (2015).

\end{thebibliography}
\end{document}